\documentclass[twocolumn,amsmath,amssymb,aps,prx]{revtex4-2}
\usepackage{graphicx}
\usepackage{dcolumn}
\usepackage{bm}
\usepackage[colorlinks=true,citecolor=blue,linkcolor=blue,urlcolor=blue]{hyperref}
\usepackage{tikz}
\usepackage{url}
\usepackage[utf8]{inputenc}
\usepackage[english]{babel}
\usepackage{amsmath}
\usepackage{amsfonts}
\usepackage{amssymb}
\usepackage{makeidx}
\usepackage{caption}
\usepackage{subcaption}
\usepackage{float}
\usepackage{natbib}
\usepackage{xcolor}

\newcommand{\be}{\begin{equation}}
\newcommand{\ee}{\end{equation}}
\newcommand{\bc}{\begin{center}}
\newcommand{\ec}{\end{center}}
\newcommand{\ben}{\begin{eqnarray}}
\newcommand{\een}{\end{eqnarray}}

\newcommand{\ket}[1]{|#1\rangle}
\newcommand{\bra}[1]{\langle #1|}
\newcommand{\braket}[2]{\langle #1|#2\rangle}
\newcommand{\op}[2]{| #1\rangle \langle#2|}

\newcommand{\id}{\mathbb{I}}
\newcommand{\Tr}{\text{Tr}}

\begin{document}

%\preprint{APS/123-QED}

\title{Entanglement detection with classical deep neural networks}

\author{Julio Ureña}
\affiliation{Instituto de F\'isica Corpuscular (IFIC), CSIC  and Universitat de València, Valencia, E-46980, Spain} 
\affiliation{Electromagnetism and Matter Physics Department, University of Granada, E-18071, Spain.}
\author{Antonio Sojo}
\affiliation{Electromagnetism and Matter Physics Department, University of Granada, E-18071, Spain.}
\author{Juani Bermejo-Vega}
\affiliation{Electromagnetism and Matter Physics Department, University of Granada, E-18071, Spain.}
\affiliation{Institute Carlos I of Theoretical and Computational Physics, University of Granada, E-18071, Spain.}
\author{Daniel Manzano}
\email[]{manzano@onsager.ugr.es}
\affiliation{Electromagnetism and Matter Physics Department, University of Granada, E-18071, Spain.}
\affiliation{Institute Carlos I of Theoretical and Computational Physics, University of Granada, E-18071, Spain.}

\begin{abstract}
Abstract:  In this study, we introduce an autonomous  method for addressing the detection and classification of quantum entanglement, a core element of quantum mechanics that has yet to be fully understood. We employ a multi-layer perceptron to effectively identify entanglement in both two- and three-qubit systems. Our technique yields impressive detection results, achieving nearly perfect accuracy for two-qubit systems and over $90\%$ accuracy for three-qubit systems. Additionally, our approach successfully categorizes three-qubit entangled states into distinct groups with a success rate of up to $77\%$. These findings indicate the potential for our method to be applied to larger systems, paving the way for advancements in quantum information processing applications.
\end{abstract}

\maketitle

\section{Introduction}

Entanglement is one of the most important features of quantum mechanics. First proposed by Einstein, Podolski, and Rosen as a pretended proof of the incompleteness of the theory \cite{einstein:pr35}, it was later considered by Schr\"odinger as {\it the characteristic trait of quantum mechanics, the one that enforces its entire departure from classical lines of thought} \cite{schrodinger:mpcps35}. Beyond its philosophical and fundamental interest, entanglement is  a crucial resource for the development of new quantum technologies, being key to techniques such as quantum teleportation \cite{bennet:prl93,pirandola:np15}, measurement-based quantum computation \cite{raussendorf:pra03,briegel:np09}, or super-dense coding \cite{bennet:prl92}. 

One problem associated with entanglement is the development of separability criteria and entanglement measures \cite{horodecki:rmp09,guhne:pr09}.  This problem is based on determining if a certain quantum system is entangled or not. Several criteria has been proposed including the celebrated Bell's inequalities \cite{bell:p64}, the Peres-Horodecki positive partial transpose criterion (PPT) \cite{peres:prl96,horodecki:pla96}, entanglement witnesses \cite{bru:jmo02, guhne:prl04}, and entropic criteria \cite{plastino:epl09,walborn:prl09}. For systems of dimension up to $6$, the PPT criteria gives sufficient and necessary conditions for a quantum state, pure or mixed, to be entangled. For systems with a higher dimension there are not known sufficient conditions. 
 
Recently, the fields of machine learning and quantum mechanics have been recently merged in the new field of {\it quantum machine learning} (QML). This connection has been made in two directions. First, quantum features can be used to enhance the learning process \cite{dunjko:prl16,dunjko:rpp18}. Second, machine learning techniques can be used to learn quantum operations and to design experiments \cite{manzano:njp09,melnikov:pnas17}. One specific line of research in these directions are quantum neural networks (QNN), meaning learning models inspired by biological systems \cite{cao:arxiv17,beer:nc20,torres:njp22}.
 
In this paper, we address the problem of separability determination by the use of a deep multilayer perceptron (MLP) \cite{rosenblatt:pr58,minsky_69}, in order to develop an autonomous method for entanglement detection.  We first test it in a solvable model, a two-qubit system, showing that it can acquire practically a $100\%$ efficiency after a small number of learning experiences and with simple topologies. We check these results in dependence with the entanglement of the system and with its purity. Furthermore, we also study the stability of the detection procedure when noise is added to the system. Finally, we apply the same method to a non-solvable model, a three qubits system, and we show that the network can reach high efficiency rates close to $100\%$ for the highest  entangled cases. For this problem we also study the performance of the network based on the different entanglement families showing that some families are easier to classify than others. 

This issue has been recently tackled in numerous studies. In Ref. \cite{roik:pra21}, the methodology employed is grounded on the utilization of entanglement witnesses with non-local measurements, applied specifically to two-qubit systems. Conversely, Ref. \cite{asif:sr23} develops a similar approach, albeit based on the results of coherence measurements. While both strategies yield noteworthy outcomes, their effectiveness is contingent upon prior knowledge about the system, delineating a clear boundary on their applicability and potential for further exploration in diverse system contexts. Finally, in Ref.  \cite{chen:qst22} an unsupervised learning algorithm is used to detect entanglement in systems up to 10 qubits by the use of genetrative adversarial networks (GANs). Moreover, by the use of a supervised learning classifiers the two qubits and two qutrit cases has been addressed in Ref \cite{lu:pra18}. Experimentally, these models have been implemented to classify quantum states without performing full tomography \cite{gao:prl18}. Finally, the complicated problem of finding separable approximations for quantum states has also been addressed by the use of neural networks \cite{girardin:prr22}.

In these works they can detect entanglement with very different efficiencies up to $97\%$ in the best case scenario for two qubits. We improve this efficiency reaching almost a $100\%$ efficiency. Besides,  we show also the capabilities of neural network to distinguish between different amounts of multipartite entanglement.  Our approach is autonomous, meaning that our network does not require any previous knowledge about the system or any kind of measurements outputs.

\section{Multilayer perceptron and the learning procedure}
\label{sec:MLP}

\begin{figure*}
\bc
\includegraphics[scale=0.15]{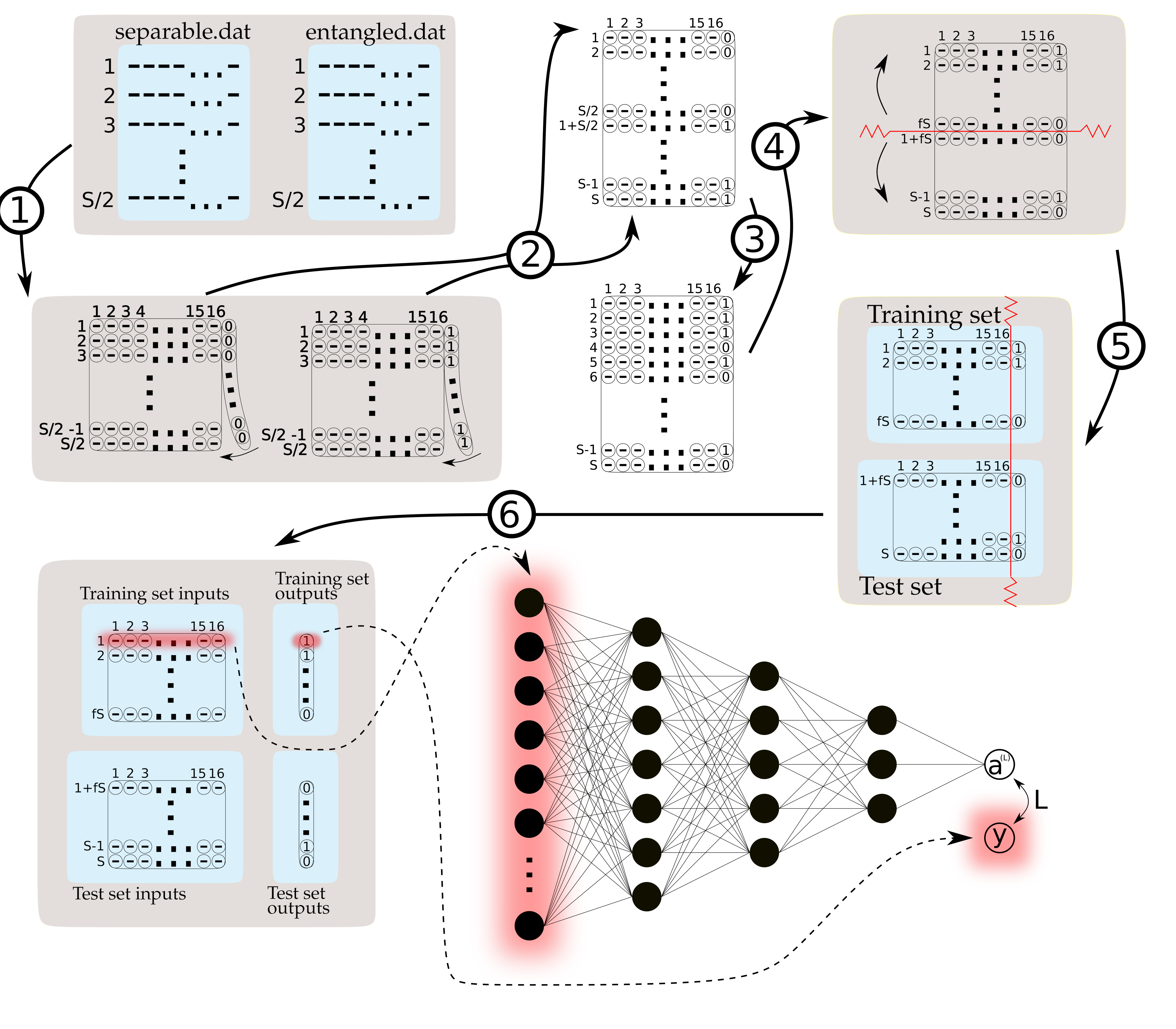}
\ec
\caption{Scheme of the data processing. Explanation in the main text. }
\label{fig:scheme}
\end{figure*}

In this section, we provide a short explanation of the MLP model used and how it is applied to our specific problem.  Our MLP is a neural network (NN) model originally based on the McCulloch-Pitts model of neurons \cite{mcculloch:bmb43} and backpropagation of the error \cite{russel_03}.  For the two-qubits case the input of the network will  be the elements of the density matrix of the state, while for the three-qubits one it will be the vector state. By doing so we ensure that the complexity of the problem is comparable in both cases.  As density matrices are Hermitian this means that for an $N$ qubits system the dimension of the matrix is $2^N\times2^N$ that corresponds to $2^{2N}$ real values that the MLP takes as input. As MLP are topologically invariant the order of the input parameters plays no major role. For the vector case the input is composed by $2^{N}$ complex values that correspond to  $2^{(N+1)}$ real independent parameters. For some specific cases we have artificially increased the input space by redundancy to improve the learning procedure.  

To analyse our network we study three figures of merit. At the end of the learning we calculate the Average Success Rate (ASR), meaning the percentage of well-classified states for the set of interest. Furthermore, to also study  the evolution of the learning procedure in binary classification problems we use the binary cross entropy (BCE) loss. If we have an output $a'$ and an ideal output $a$ the BCE is defined as 

\be
BCE(a,a')=-\left( a \, \log(a') + (1-a) \, \log(1-a')  \right).
\label{eq:bce}
\ee
Finally, in Section \ref{sec:3qubits} we also studied the problem of classifying four entanglement families. In this case, the output layer consists of four neurons with activations  $a_i$,  $i\in\left\{1,\,2,\,3,\,4\right\}$, each of them corresponding to one of the families. For this specific case the readout of the MLP is not the activation of the neurons but the softmax function of these activations defined as 

\be
S(a_i)=\frac{\exp (a_i) }{\sum_{j=1}^4 \exp (a_j)}.
\ee
This can be considered as a probability distribution defined over the four entanglement families. The considered loss function for this problem will be the cross entropy between the predicted probability distribution $\left\{S(a_i) \right\}_i$ and the desired one that is just $\delta_{ii'}$, being $i'$ the correct classification and $\delta$ the Kronecker-delta function. Therefore, we define the Categorical Cross Entropy (CCE) as 

\be
CCE(a_i,i') = -\sum_j \delta_{ji'}\, \log \left( S(a_j) \right) = -\log\left( S(a_{i'})  \right),
\label{eq:cce}
\ee
that depends only on the softmax function of the neuron which is associated to the correct classification.

The training procedure is organised in epochs of training.  Although more than one estimations of the loss function gradient may be performed during one epoch, every sample of the training set contributes on average once to the cited estimations.  Datasets have a size $S$ that is generally divided into $S/2$ separable and $S/2$ entangled states. The batch size, $M$ is the number of samples that are processed before the MLP is updated once.  Unless stated otherwise, we assume  $M=40$. The parameter $f$ is the fraction of the whole dataset that is used for training and it is set to $f=0.8$ except  where stated otherwise. As $M$ takes an integer value between one and the number of samples in the training set, $f\cdot S$, one epoch takes $(f\cdot S)/M$ updates of the trainable parameters.  The datasets are generated randomly (see appendix \ref{ap:generation}).

Once the density matrices for both separables and entangled states are generated and stored in files, the data should be prepared to be computed by the MLP. The data processing is sketched in Fig \ref{fig:scheme}. It is divided into six steps. {\bf Step one:} The separable and entangled density matrices (or state vectors in the three-qubits case) are read from the files and transformed into real vectors of size $2^{2N}$, generating two arrays of size $\left(2^{2N}+C \right) \times S/2$ where $C$ is one for the case of a binary categorical classification, meaning that we add the value $0/1$ to classify separable/entangled states. For the three-qubits classification problem $C=4$ as we add to the input vector a new vector of dimension 4 with each element determined by $\delta_{ji'}$, being $i'$ the family of the state and $j=1,\;2,\;3,\;4$ each vector element. {\bf Step two:} Both arrays are stacked, giving rise to the whole dataset of $S$ samples. {\bf Step three:} The array is randomly shuffled to mix the separable and entangled density matrices.  {\bf Step four and five:} The dataset is split into the training set, of size $f\cdot S/2$, and the test set of size $(1-f)\cdot S/2$. {\bf Step six:}  Both the training and test set are split up into the input set (density matrices/state vectors) and output set (binary label for binary classification cases and four dimension vectors for 3-qubits categorical classification). The rows of the input set are fed into the input layer of the MLP. The output is used, together with the true label $a$ to calculate the BCE by Eq. (\ref{eq:bce}). 

The activation function for the hidden layers is set to the Rectified Linear Unit (ReLU) \cite{nair:proc10,du:m22}, the output layer has a sigmoidal activation function  \cite{russel_03}. To backpropagate the error and optimise the network we use both the Adam Optimization Algorithm \cite{diederik:proc15} and the Root Mean Squared Propagation (RMSProp) \cite{riedmiller:proc93} as indicated in the caption of each figure. The simulations have been performed by the use of Python 3.8.10 and the libraries  NumPy 1.21.4, Pandas 1.3.4, Tensorflow 2.7.0, Keras 2.7.0, and Scipy 1.7.2. The initial values of weights and biases of the network have been stablished by the uniform Glorot method of the Keras library \cite{glorot:proc10}.

\section{Entanglement detection for two qubits systems}
\label{sec:2qubits}

\begin{figure}
\bc
\includegraphics[scale=0.6]{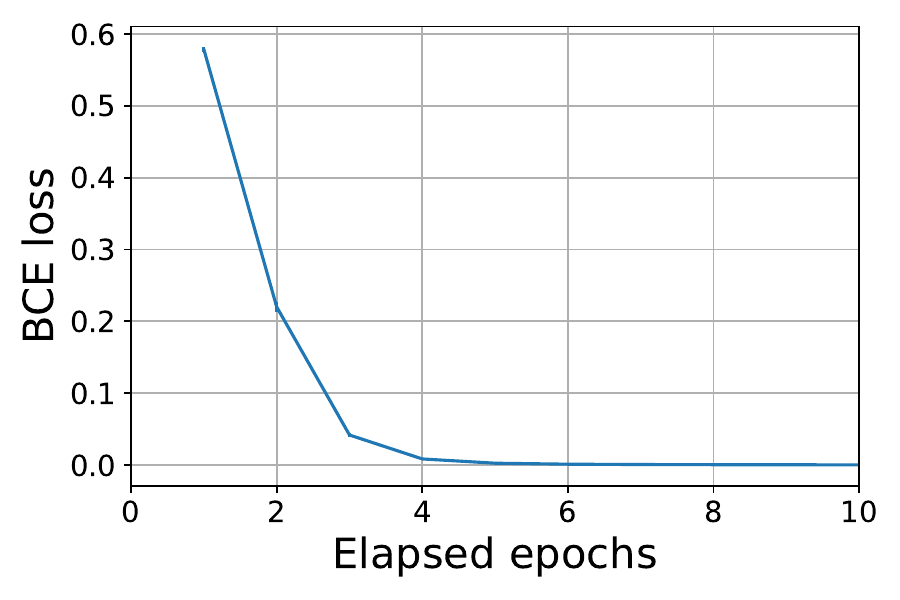}
\ec
\caption{ Differentiation of maximally entangled states from separable states for pure states. The curve represents the BCE loss as a function of the elapsed epochs of the learning process measured over the samples of the training set. The curve is averaged over 100 simulations and belong to MLP with a topology $\langle 16:8:1\rangle$. RMSprop optimizer was used and the dataset size is $S= 2 \cdot10^4$.}
\label{fig:fig1}
\end{figure}

The first problem we have studied is the detection of entanglement in an analytically solvable case, a two-qubit system. For this case necessary and sufficient separability conditions are given by the PPT separability criteria \cite{peres:prl96,horodecki:pla96,horodecki:rmp09}. First of all we study the capability of the network in order to classify totally separable states from maximally entangled ones. In appendix \ref{ap:generation}, there is a description of the method used to generate the datasets.  For this problem the network achieves a $100\%$ efficiency even with a simple topology, with only two hidden layers, and in a small number of epochs. 

In Figure \ref{fig:fig1}, we can see the evolution of the binary cross entropy (BCE) loss as a function of the number of elapsed epochs of training.
 After a very small number of epochs, the network is able to classify the separable and maximally entangled states with practically $100\%$ efficiency evaluated over the training set. Even if this is a solvable model this result is remarkable. The PPT criteria is a complex procedure that involves both partially transposing and eigenvalue calculation. The fact that a neural network can learn an equivalent procedure in less than five epochs highlights the potential of MLPs in the problem of entanglement detection.

Furthermore, the network can also be trained to detect entanglement for non-maximally entangled states. As a measure of the amount of entanglement we have used the negativity. For a general state $\rho$ its negativity is defined as 

\be
{\cal N} (\rho) = \frac{\left|\left| \rho^{T_1}  \right|\right| - 1 }{2},
\ee
where $\rho^{T_1}$ represents the partial transpose of the density matrix with respect to the first qubit and $\left|\left| A \right|\right|\equiv \Tr \sqrt{A^{\dagger} A}$ is the trace norm. The maximum value of negativity for a two-qubits case is $0.5$, meaning that the system is fully entangled. 

\begin{figure*}
\centering
\begin{subfigure}{.49\textwidth}
  \centering
  \includegraphics[width=\textwidth]{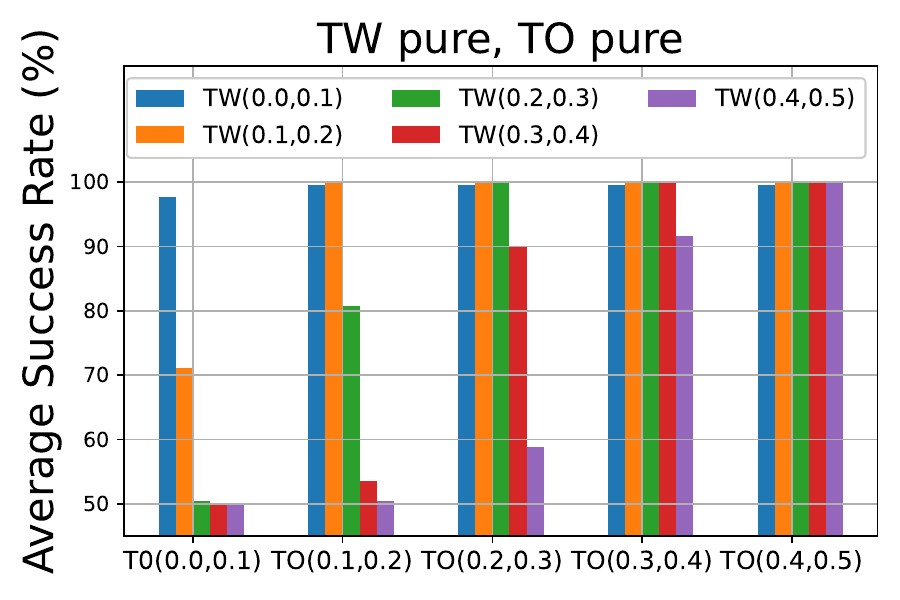}
\end{subfigure}
\begin{subfigure}{.49\textwidth}
  \centering
  \includegraphics[width=\textwidth]{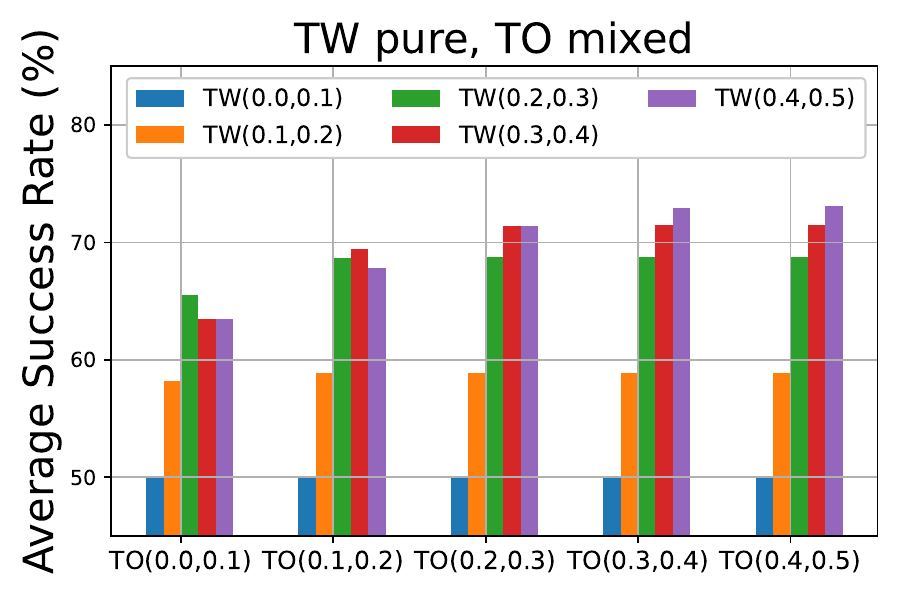}
\end{subfigure}

\begin{subfigure}{.49\textwidth}
  \centering
  \includegraphics[width=\textwidth]{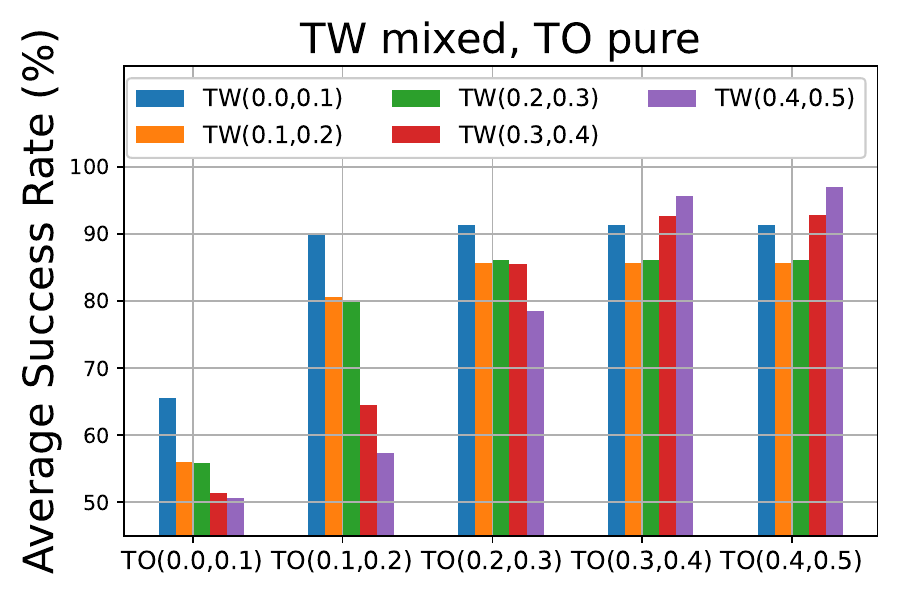}
\end{subfigure}
\begin{subfigure}{.49\textwidth}
  \centering
  \includegraphics[width=\textwidth]{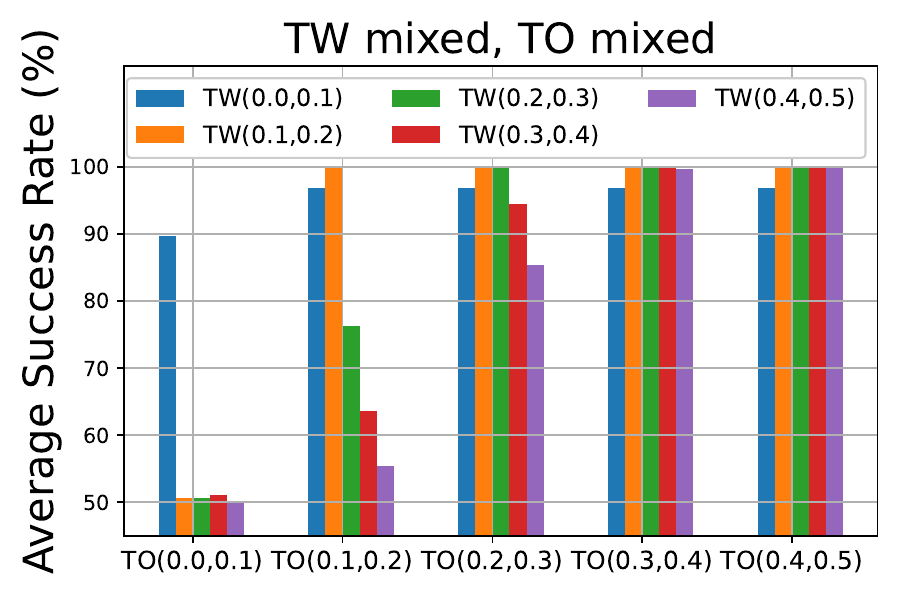}
\end{subfigure}

\caption{ASR of MLP's that resulted from the training with pure (up) and mixed (down) states, when tested over datasets containing pure (left) and mixed (right) entangled states with different negativities. TW(a,b) stands for ’trained with’, while TO(a,b) stands for ’tested on’, while the numbers on the parenthesis represents the negativity bounds of the set. The success rate for TW and TO datasets that belong to different negativity subinterval are averaged over the whole datasets, whereas the ASR's of MLPs which have been TW and TO datasets which belong to the same negativity interval are averaged over the original test set (20\% of the data). The results are averaged over 10 simulations. The topologies used  for pure states are:  ${\cal N}\in(0.0,\, 0.1)\to \langle 256:128:16:1 \rangle$,  $(0.1,\, 0.2)\to \langle 128:16:1\rangle$, $(0.2,\, 0.3)\to \langle 64:16:1 \rangle$ , $(0.3,\, 0.4)\to \langle 32:4:1 \rangle$,  $(0.4,\, 0.5)\to \langle 16:4:1 \rangle$. 
For mixed states the topologies are:  ${\cal N}\in(0.0,\, 0.1)\to \langle 256:128:16:1 \rangle$, $(0.1,\, 0.2)\to \langle 128:16:1 \rangle$, $(0.2,\, 0.3)\to\langle 64:8:1 \rangle$, $(0.3,\, 0.4)\to \langle 16:4:1 \rangle$,  $(0.4,\, 0.5)\to \langle 16:1 \rangle$. The datasets sizes are $S=2\cdot 10^4$.  }
\label{fig:fig2}
\end{figure*}

In this analysis we classify random quantum states in sets with different amounts of entanglement. These sets allow us to study the performance of the networks depending on the negativity of the entangled states of the training set in comparison with the test set. The results are displayed in Fig \ref{fig:fig2}. In this plot we can see the Average Success Rate (ASR) for the test set, after the network has been fully trained, when the system is trained with (TW) sets of different negativity values, and it is tested on (TO) the different  test sets. We have ordered the sets in $0.1$  negativity width intervals, meaning that there are five negativity intervals. The analysis has been performed by training and testing the MLP for both pure and mixed states. For this problem the optimal topology of the network strongly depends on the negativity of the training sample. Samples with lower negativities require deeper and bigger networks to be properly classified.  For all cases we have profoundly studied different topologies and selected the simplest one that achieves the maximum efficiency.

The first conclusion that can be drawn from Fig. \ref{fig:fig2} is that networks trained with pure/mixed states are optimal for detecting entanglement only for pure/mixed states. The entanglement signatures in the density matrix strongly depend on the purity of the states, and the learning procedure is affected by this. However, a network trained with mixed states performs slightly better when tested on pure states than the other way around. Another conclusion is that only networks trained with states of low negativity are capable of detecting the entanglement of these same states. Furthermore, these networks perform well with states with higher negativity. However, this does not happen in the other direction, as networks trained with highly entangled states lose efficiency when applied to less entangled sets. These results indicate that networks should be trained for worst-case scenarios so they can perform well in any case.

\begin{figure}
\bc
\includegraphics[scale=0.5]{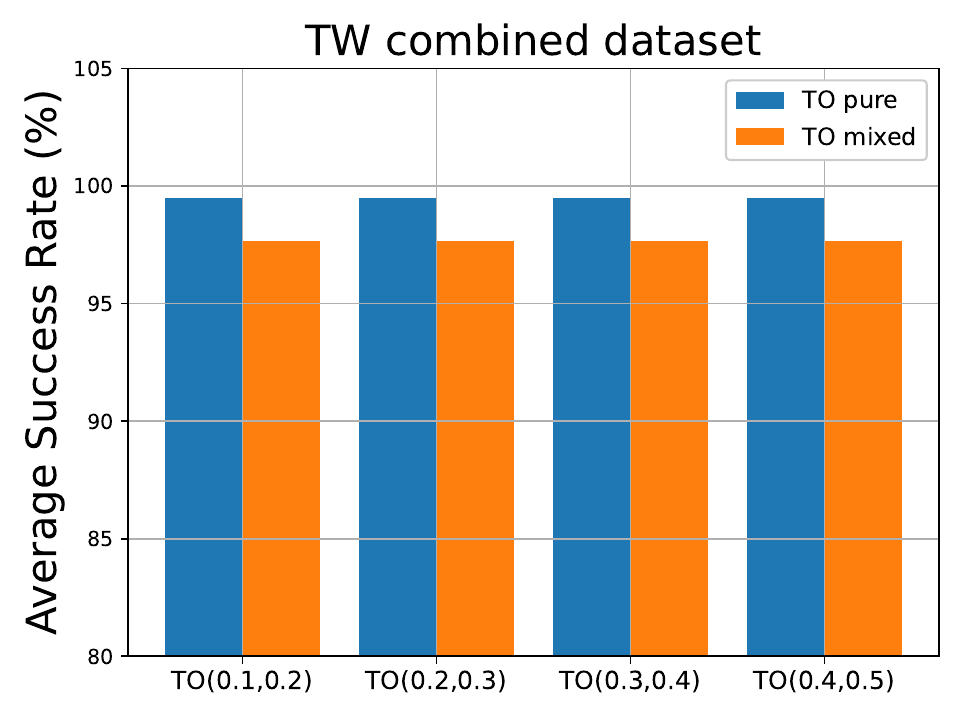}
\ec
\caption{ ASR of the MLPs that resulted from training with a mixture of minimally entangled $({\cal N} \in (0, 0,1)$) pure and mixed states when applied to both pure and mixed test sets with different negativities. The activation function in the hidden layers is ReLU, and the used optimizer is RMSprop. Efficiency is averaged over ten simulations.  The MLP architecture is $\langle 256 : 128 : 16 \rangle$. The datasets sizes for this case is $S=4\cdot10^4$.}
\label{fig:epsilon}
\end{figure}

Based on these results we can infer that a network trained with both pure and mixed states with low entanglement may be the most general entanglement detector for arbitrary states (mixed and pure with arbitrary negativity). To test this hypothesis we have trained a network with a dataset composed by a shuffling of pure and mixed states with negativity ${\cal N}\in (0,\,0.1)$. After the training procedure we tested the network in training sets with both pure and mixed states with different negativities. In this case, as it is shown in Fig.  \ref{fig:epsilon} , regardless the test set we obtain an average success rate higher than 97\%

%%%%%%%%%%%%%%%%%%%%%%%%%%%%%%%%%%%%%%%%%%%%%%%%%%%%%%%%%

To further analyse the performance of the network in the boundaries between separable and entangled states, we have studied two specific two-qubits families. First, we have trained the network with systems of arbitrary negativity and  checked the probability of determining that a certain state is entangled for states 

\be
\ket{\psi_\epsilon} = \frac{\ket{\psi_\text{sep}} + \epsilon \ket{\psi_\text{Bell} }}
{\sqrt{1 + \left| \epsilon \right|^2 + 2 \text{Re}\left\{  \epsilon   \; \braket{\psi_\text{sep}}{\psi_\text{Bell} \right\} } }},
\label{eq:epsilon}
\ee
where in general $\epsilon\in\mathbb{C}$ but we have studied only the case of $\epsilon\in[0,1]$, $\ket{\psi_\text{sep}}$ are bipartite separable states, and $ \ket{\psi_\text{Bell}}$ are maximally entangled states. These states are separable only in the limit $\epsilon=0$ and entangled otherwise.

The purpose of studying these types of states is to evaluate the network's robustness to noise. As the volume of separable states is much smaller than that of entangled states, especially for pure systems \cite{zyczkowski:pra98}, it is expected that if a separable system is subject to static noise, it will become  a random state and it will be, with a high probability, entangled as the density of entangled states is exponentially higher than the one of separable states. Therefore, it would be desirable if the network can correctly classify states that are close to being separable as separable. However, for some contexts this could also be considered a failure in entanglement detection. The parameter $\epsilon$ at which the network detects entanglement would vary depending on the specific states being analysed, making this case interesting to study and classify.

\begin{figure*}
\bc
\includegraphics[scale=0.5]{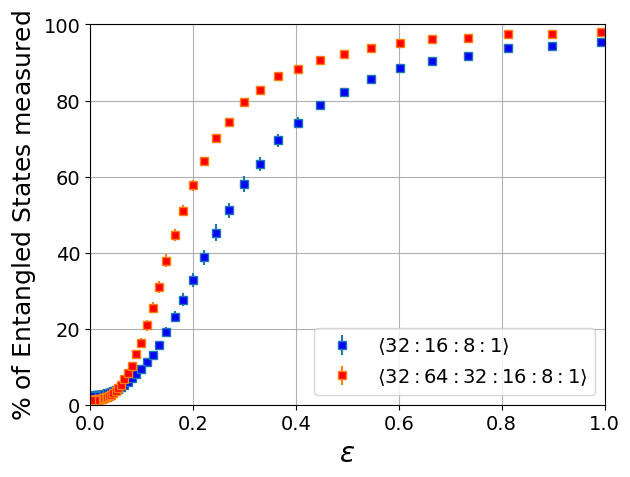} \hspace{1cm}
\includegraphics[scale=0.5]{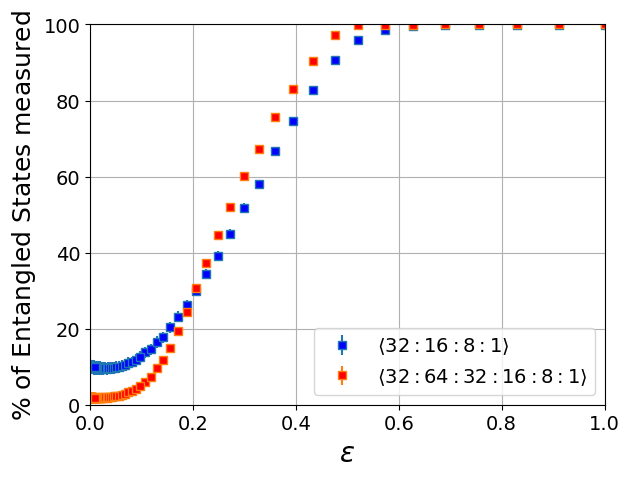}
\ec
\caption{Probability of determining that a certain state is entangled as a function of the parameter $\epsilon$ for states of the form (\ref{eq:epsilon}) for pure (left) and mixed (right) states. }
\label{fig:epsilon2}
\end{figure*}

The results are presented in Figure \ref{fig:epsilon2} for both pure and mixed states (see Appendix for details about the generation of states). Interestingly, for mixed states and a small network with only two hidden layers, the classification performance is not robust under the presence of small noise, as it classifies up to $15\%$ of states as entangled for small $\epsilon$ values. However, if we increase the network depth to four hidden layers, the learning becomes more robust, and it classifies almost all states as separable if $\epsilon<0.1$. For higher values of $\epsilon$, both networks behave similarly. For pure states the behaviour of both networks is similar but there are qualitatively differences, being the deeper network more efficient. This result suggests that, although the success probability of deep and non-deep networks is similar, deep networks are able to capture more entanglement features during the learning process. Hence, depending on the purpose of the network it would be more interesting to design it with a specific topology. 

\begin{figure}[h]
\bc
\includegraphics[scale=0.5]{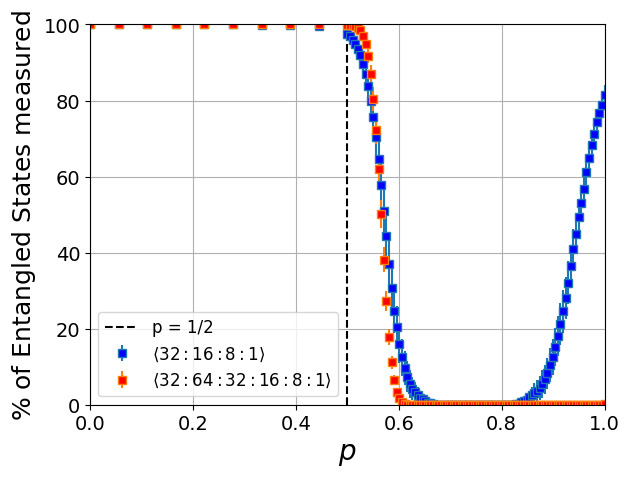}
\ec
\caption{Probability of determining that a certain state is entangled as a function of the parameter $p$ for states of the form (\ref{eq:werner}).}
\label{fig:werner}
\end{figure}

A similar result is obtained by studying Werner states in the form \cite{werner:pra89}

\be
\rho_W= \frac{p}{3} \id + (1-\frac{4p}{3}) \op{\psi_\text{Bell}}{\psi_\text{Bell}}
\label{eq:werner}
\ee
where $p\in[0,1]$, $\id$ represents the maximally mixed state, and $\ket{\psi_\text{Bell}}$ is a maximally entangled state (see Appendix). These kind of states  are entangled if, and only if, $p<\frac{1}{2}$. In this case, as it is shown if Figure \ref{fig:werner}, both networks overestimate the presence of entanglement, detecting more than half of the states as entangled up to $p\sim 0.55$.  A more interesting feature arises for values of $p$ higher than $0.8$. In this case, the network with two hidden layers starts detecting again the states as entangled. On the other hand, the network with four hidden layers makes a correct classification also in this case.  This indicates that the deeper networks are more able to identify pure entanglement properties of the systems while the smaller ones can be tricked by other properties of the dataset as the rank of the density matrices. Interestingly, in Refs \cite{roik:pra21,asif:sr23} a similar result is obtained for a very different neural network model and training procedure.

\section{Entanglement detection and classification for three qubits systems}
\label{sec:3qubits}

It is well-known that three-qubit systems exhibit much more complex behaviour with respect to their separability properties. It has been proven that there are six possible entanglement classes, meaning six types of states that can be connected by stochastic local operations and classical communication (SLOCC). These are the separable, bipartite entangled (BE), Greenberger-Horne-Zeilinger (GHZ), and W classes \cite{greenberger_89,dur:pra00}. The bipartite class can be further divided into three classes, depending on the way in which the bipartition is performed. In our case, where the three qubits are identical, we consider all three types of bipartite entanglement to belong to the same entanglement family. To keep the complexity of the problem comparable to the two-qubit case, we have only worked with pure states in this section. The MLP input consists of the 16 real parameters corresponding to the elements of the state vector.

\begin{figure*}
\begin{subfigure}{.45\textwidth}
  \centering
  \includegraphics[width=\textwidth]{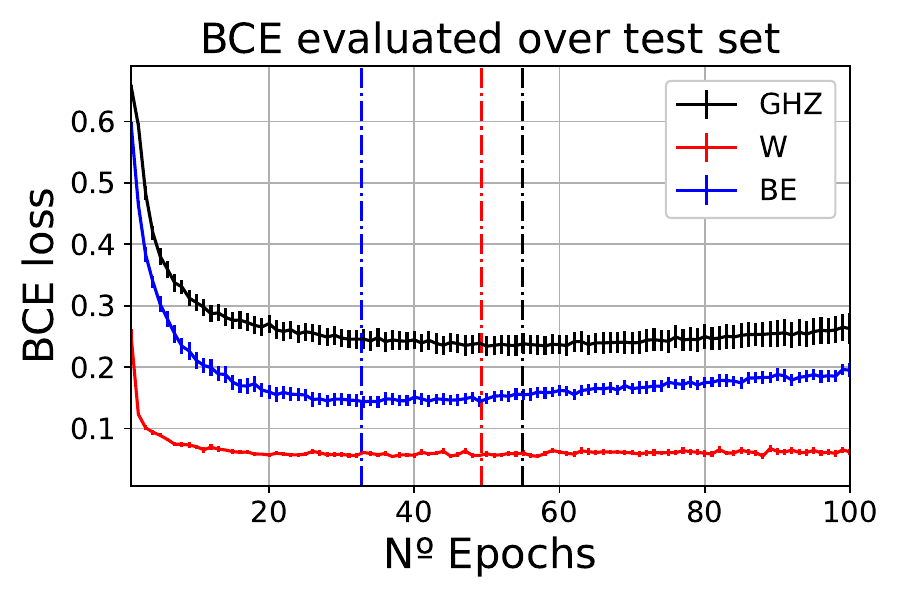}
\end{subfigure}
\begin{subfigure}{.45\textwidth}
  \centering
  \includegraphics[width=\textwidth]{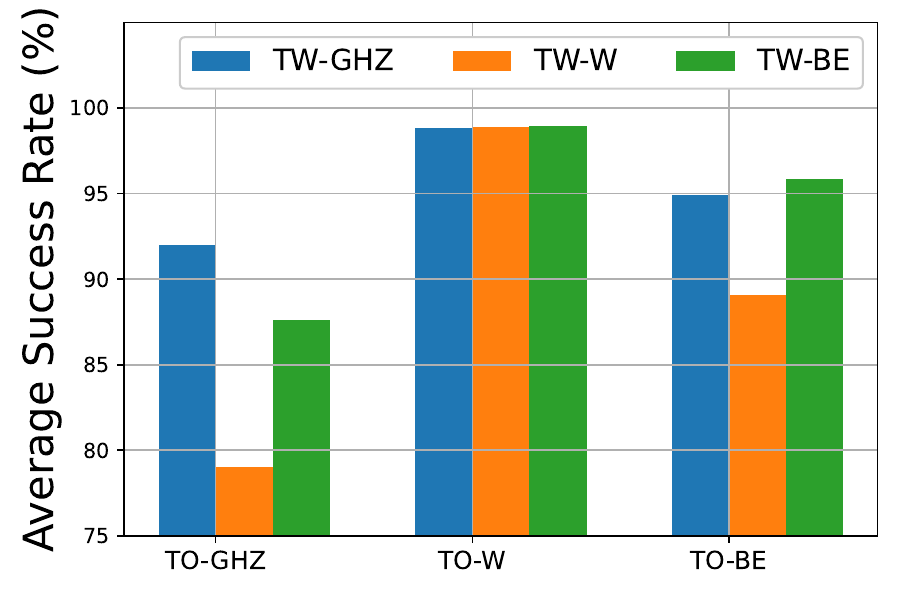}
\end{subfigure}
\caption{Left: BCE loss as a function of the epochs of learning for the three entanglement families. The architecture of the network is $\langle 16:512:128:32:1 \rangle$. The fraction of states in the dataset taken for training is $f=0.75$ and the BCE is evaluated only in the test set. Each curve is averaged over 10 simulations. The vertical lines indicate the epochs in which the best configuration is reached. The datasets sizes are $S=2\cdot 10^5$.}
\label{fig:3qubits}
\end{figure*}

First, we checked the learning rate of the MLP for each of the three families. Figure \ref{fig:3qubits} (left) shows the BCE loss as a function of the learning procedure for each family. The vertical coloured lines represent the moment when the best configuration is achieved, and the MLP starts to suffer from overfitting. We can see that each family has a different learning speed, as well as a different maximum achievable efficiency. The W states are the easiest to classify, reaching a BCE loss below $0.1$ after just 10 epochs. They are also the family that achieves the highest detection efficiency. The worst learning scenario occurs for the GHZ states, where the BCE loss cannot go below $0.2$. This result is very interesting and indicates a relation between the amount of tripartite entanglement and the rate of entanglement detection. It is known that the W family contains the only states with entanglement that can be considered both tripartite and bipartite, while GHZ states can be considered only tripartite entangled \cite{dur:pra00}. From this result, we can conclude that tripartite entanglement is the hardest to detect with our MLP, bipartite entanglement is easier, and W states are the easiest as they contain both tripartite and bipartite entanglement.

After training with the selected entanglement class, in each case, we evaluated the performance of the MLP on all three classes. The results are shown in Figure \ref{fig:3qubits} (right). The MLP trained with W states and applied to GHZ states had the worst performance, followed by the situation where the training was performed with BE states and the network was again applied to the GHZ family. This supports our claim that bipartite entanglement is easier to learn than tripartite entanglement, and when networks are trained with states containing bipartite entanglement, they perform poorly when faced with tripartite entanglement. On the other hand, the best performance was obtained when the network was applied to the W states regardless of the training set. This may be due to the presence of both tripartite and bipartite entanglement in this family. Additionally, we observed that when the MLP was trained with W states, it performed poorly when applied to any other family.

\begin{figure*}
\begin{subfigure}{.45\textwidth}
  \centering
  \includegraphics[width=\textwidth]{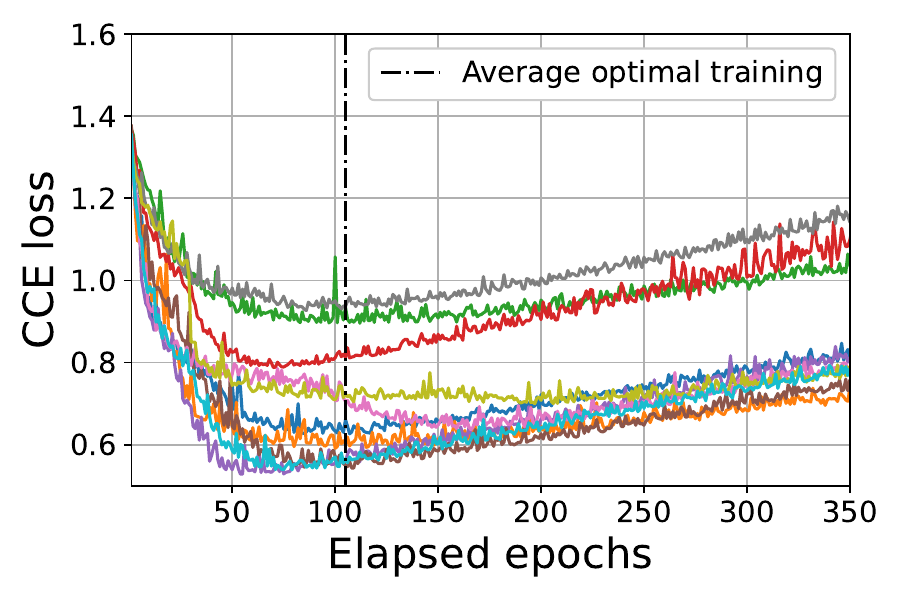}
\end{subfigure}
\begin{subfigure}{.45\textwidth}
  \centering
  \includegraphics[width=\textwidth]{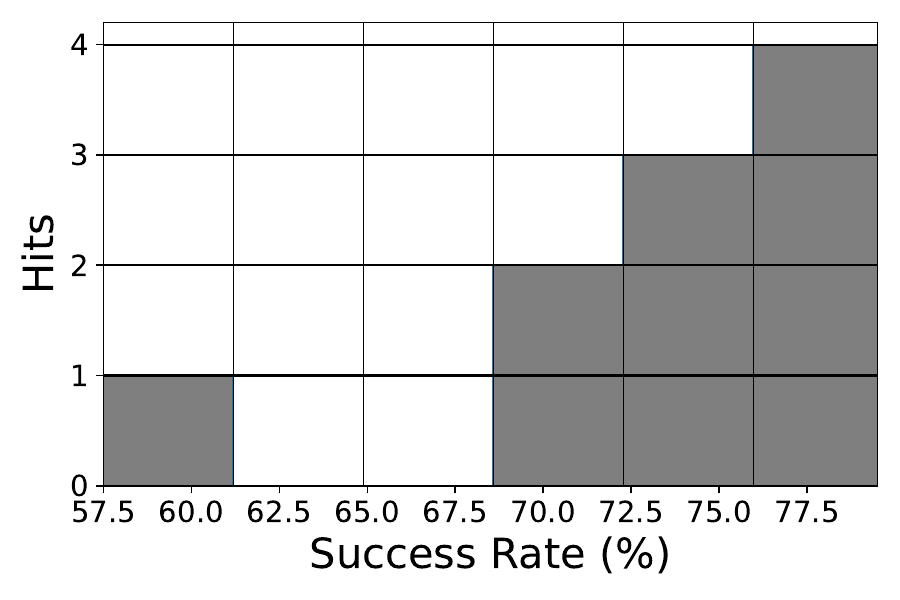}
\end{subfigure}
\caption{Left: CCE of the categorical classification for 10 runs with different initial conditions over the same dataset. The vertical line marks the average number of epoch corresponding to the optimal training determined by an Early Stop algorithm. In these cases, the batch size is $M = 1000$ and $f = 0.75$. Right: Histogram of all 10 ASR samples which result from each simulation. The topology of the network is $\left<16:512:128:32:4  \right>$. The dataset size is $S=4\cdot 10^5$.  }
\label{fig:3qubitsCCE}
\end{figure*}

Finally, we tested the MLP's ability to classify the states into four possible families: bipartite entangled, W, GHZ, and separable. To do this, we used a network with four output neurons as explained in Section \ref{sec:MLP}. For this problem, the initial conditions of the network, meaning the initial weights and biases,  were very important, as shown in Figure \ref{fig:3qubitsCCE} (left). We plotted the CCE as a function of the number of epochs for 10 different runs over the same dataset. Each run differed in the initialization of the network weights and the ordering of the dataset elements. As can be seen, different initial conditions led to different behavior with respect to the speed of learning, the final efficiency, and overfitting.

For these 10 runs the best final success rate achieved is $79.7\%$ and the average one is $73.2\%$. In Figure \ref{fig:3qubitsCCE} (right) we can observe the number of runs that have lead to each different success rate. The wort case, $57.5\%$ seems an unlikely event while most of runs lead to a final efficiency above $70\%$. We may also remark that in this case the probability of correctly classifying a state by a purely random procedure is $25\%$ instead than $50\%$ as is in the binary classification. 

\section{Conclusions}
In this study, we have showcased the potential of deep learning algorithms to effectively address entanglement detection and classification challenges. Our findings are striking, as the network attains up to $100\%$ efficiency in two-qubit scenarios and over $90\%$ efficiency in three-qubit situations. Additionally, we identified a strong relationship between entanglement and purity, with networks trained on pure states underperforming when presented with mixed states and vice versa.

Moreover, deep networks display resilience to minor noise and can identify entanglement characteristics that allow them to excel when working with well-established quantum families like Werner states. In three-qubit instances, the network can pinpoint the entanglement family of a state with over $77\%$ precision. Our research introduces a novel approach to detecting and classifying entanglement that bypasses the need for specific criteria or witnesses tailored to particular dimensions.

This study paves the way for further exploration in several areas. Firstly, investigating the impact of state properties such as purity on neural network performance could lead to improved detection algorithms and a deeper understanding of the interplay between various quantum properties. The techniques proposed here can also be adapted for other quantum information tasks, such as state comparison. Lastly, it would be valuable to investigate the creation of a quantum neural network capable of executing the same tasks, which could open up a myriad of applications and serve as a benchmark problem for both classical and quantum neural networks.

\section{Declarations}

\subsection{Ethical Approval and Consent to participate}

Not applicable. 

\subsection{Consent for publication}

All authors consent for the publication of this paper. 

\subsection{Availability of supporting data}

The data supporting this research is available upon request to the corresponding author D. Manzano. 

\subsection{Competing interests/Authors' contributions}

There are no competing interests. 

D.Manzano and J.J. Bermejo-Vega had the idea and developed the project. J. Ureña and A. Sojo generated the data, performed all simulations, and made the plots. D. Manzano and J.J. Bermejo-Vega supervised and extract the conclusions. D. Manzano wrote the manuscript and all authors reviewed and improved it.

\subsection{Funding}

We want to acknowledge funding from the FEDER/Junta de Andaluc\'ia program A.FQM.752.UGR20, and project PID2021-128970OA-I00 funded by MCIN/AEI/ 10.13039/501100011033 and, by ``ERDF A way of making Europe'', by the ``European Union'', the Ministry of Economic Affairs and Digital Transformation of the Spanish Government through the QUANTUM ENIA project call – Quantum Spain project, and by the European Union through the Recovery, Transformation and Resilience Plan – NextGenerationEU within the framework of the Digital Spain 2026 Agenda

\newpage
\appendix

\section{ Generation of random quantum states}
\label{ap:generation}

To begin generating random states, our first step is to create a random one-qubit unitary operation. For this purpose, we rely on a Euler's angles parametrisation \cite{nielsen_00}.

\be
U=\left(  
\begin{array}{ccc}
e^{i(\theta_1- \frac{\theta_2}{2} - \frac{\theta_3}{3})} \cos \left(\frac{\theta4}{2} \right) & &  -e^{i(\theta_1- \frac{\theta_2}{2} + \frac{\theta_3}{3})} \sin \left(\frac{\theta4}{2} \right) \\
e^{i(\theta_1+ \frac{\theta_2}{2} - \frac{\theta_3}{3})} \sin \left(\frac{\theta4}{2} \right) & &  e^{i(\theta_1+ \frac{\theta_2}{2} + \frac{\theta_3}{3})} \cos \left(\frac{\theta4}{2} \right)
\end{array}
\right),
\ee
with $\theta_i\in [0,2\,\pi[$. To generate a random unitary operation, $U^\text{rand}$ we just sample these angles from a uniform distribution. As local unitaries preserve the amount of entanglement we use them to generate random states while keeping their negativity values.

\subsection{Two qubits}

\subsubsection{Separable states}

{\bf Pure: }
We take as a starting point the state $\ket{\psi_\text{sep}^0}= \ket{0}\otimes\ket{0}$ and we apply to it an operator in the form  $U=U_{1}^{rand} \otimes U_{2}^{rand}$. The result would be 

\be
\ket{\psi_\text{sep}^{\text{rand}}}= U \psi_{\text{sep}^0} =  U_{1}^{rand}  \ket{0} \otimes U_{2}^{rand}  \ket{0}.
\label{eq:random_sep}
\ee
{\bf Mixed: }
By definition, any two system separable mixed state is written as a convex combination of tensor product of density matrices of each system:
\begin{equation}
    \rho = \sum_{i=1}^L p_i \rho^{(1)}_i \otimes \rho^{(2)}_i
\end{equation}
with $\sum_{i} p_i = 1$ and $0\le p_i \le 1$. To generate random separable mixed states with just need to generate $L$ normalized coefficients $p_i$ and $L$ pairs of $1$ qubit density matrices. We draw $L$ numbers $\tilde{p}_i$  from an uniform distribution between $0$ and $1$ and then they are normalized  to obtain each $p_i = \tilde{p}_i/\sum_i\tilde{p}_i$. The parameter $L$ allows us to control the matrix rank of our datasets. We use $L\in[2,7]$  and we post select the data to ensure that the datasets are uniformly distributed between rank 2 and 4. 

\subsubsection{Maximally entangled states: }

{\bf Pure:}
We apply a similar procedure than in the separable case but starting with $\ket{\psi_+}=\left( \ket{00} +\ket{11} \right)/\sqrt{2}$. Therefore, the random states are

\be
\ket{\psi_\text{Bell}^{\text{rand}} }= \left( U_{1}^{rand}  \otimes U_{1}^{rand}   \right)  \ket{\psi_+}.
\label{eq:random_bell}
\ee
For this case we do not generate mixed states because for two qubits all maximum entangled states are pure.

\subsubsection{Non-maximally entangled states }
{\bf Pure:}
It is more complicated to generate random states with an arbitrary value of the negativity. To do so, we generate random two-qubit states and postselect the dataset by measuring the negativity. We start with an arbitrary basis of the two-qubits Hilbert space $\left\{ \ket{u_i}, \; i=1,\dots,4 \right\}$. Any state can be expanded in this basis as 

\be
\ket{\psi} = \sum_{j=1}^4 r_j e^{i \phi_j} \ket{u_j},
\ee
with $r_j\geq 0$, $\phi\in[0,2\pi[$, and $\sum_{j=1}^4 r_j^2=1$. Therefore, to generate random two-qubits systems we sample eight real numbers $r_j\in[0,1]$ and $\phi_j\in[0,2\pi]$. The normalized random state would be

\be
    \ket{\psi^{\text{rand}}} = \frac{1}{\left(\sum _{k=1} ^4 r_k ^2 \right)^{1/2}} \sum _{j=1} ^4 r_j e^{i\phi _j} \ket{u_j}. 
\ee
To prepare sets of states with different negativity values we have just generated enough random states to sort them. 

\vspace{0.2cm}
\noindent
{\bf Mixed: }
In this case, mixed states are calculated by a linear combination of pure states created by the previous method. Again, for each mixed state we use a number of pure states $L\in[2,7]$ distributed in a way that ensure that the datasets are uniformly distributed between rank 2 and 4.

\subsubsection{$\epsilon$-states} 

{\bf Pure:}
For the $\epsilon$-states of section \ref{sec:2qubits} with a fixed value of $\epsilon\in[0,1]$ we generate random states in the form 

\be
\ket{\psi_\epsilon^{\text{rand}}} = \frac{\ket{\psi_{\text{sep}}^{\text{rand}}} + \epsilon \ket{\psi_\text{Bell}^{\text{rand}} }}{\sqrt{1 + \left| \epsilon \right|^2 + 2 \text{Re}\left\{  \epsilon   \;  \braket{\psi_{\text{  sep}}^{\text{rand}}  }{\psi_\text{Bell}^{\text{rand}}}  \right\} }},
\ee
with $\ket{\psi_\text{sep}^{\text{rand}}}$ and $ \ket{\psi_\text{Bell}^{\text{rand}}}$ generated by the use of Eqs. (\ref{eq:random_sep}) and (\ref{eq:random_bell}) respectively. 

\vspace{0.2cm}
\noindent
{\bf Mixed: }
To define the mixed $\epsilon$-states, we use the previous construction to compute the next density operator as the convex combination:

\begin{equation}
    \rho(\epsilon) = (1-\epsilon) \ket{\psi_{\text{sep}}^{\text{rand}}}  \bra{\psi_{\text{sep}}^{\text{rand}}} + \epsilon \ket{\psi_\text{Bell}^{\text{rand}}} \bra{\psi_\text{Bell}^{\text{rand}}},
\end{equation}
and as we are working with $\epsilon\in[0,1]$ the result is a positive density matrix (this would not work for other choices of $\epsilon)$.

\subsubsection{Werner states:}

The Werner states of section \ref{sec:2qubits} are mixed by definition and are prepared by  

\be
\rho_W(p)=\frac{p}{3} \id + \left( 1-\frac{4p}{3} \right) \left( U^{\phantom{\dagger}}_1\otimes U^{\phantom{\dagger}}_2\right) \ket{\psi_-} \bra{\psi_-} \left( U^\dagger_1\otimes U^\dagger_2 \right),
\ee
where $U_1$ and $U_2$ are single qubit random rotations and $p\in [0,1]$.

\subsection{Three qubits}

When working with three qubits, there are six equivalence classes that need to be generated. For three qubits, we only consider pure states. 

\subsubsection{Separable}
Separable states are prepared by applying three one-qubit operators to a fixed initial state 

\be
\ket{\psi_\text{sep}^{\text{rand}}}_3=  U_{1}^{rand}  \ket{0} \otimes U_{2}^{rand}  \ket{0} \otimes U_{3}^{rand}\ket{0}.
\label{eq:random_sep_3}
\ee

\subsubsection{Bipartite entangled states}
To generate bipartite entangled states we first select which qubits are going to be separated by sampling a random number between 1 and 3. This qubit is set to $\ket{0}$, the other two qubits are set to the maximally entangled state $\ket{\psi_+}$. We then apply a three qubits random unitary operator. For instance, is we select the first qubit to be separated a random state would be calculated as: 

\be
\ket{\psi_{\text{BE}}} = U^{\text{rand}}_1 \ket{0} \otimes \left( U^{\text{rand}}_2 \otimes U^{\text{rand}}_3  \right) \ket{\psi_+},
\ee
the extension for the other two cases is straightforward.

\subsubsection{GHZ-states:} 

We use the following parametrisation

\be
\ket{\Psi_{\text{GHZ}}} = \sqrt{K_{GHZ}} \left( \cos(\delta)  \ket{0} \ket{0} \ket{0} + \sin(\delta) e^{i\phi}  \ket{\varphi_A} \ket{\varphi_B} \ket{\varphi_C} \right),
\ee
with

\ben
 \ket{\varphi_A} &=& \cos(\alpha) \ket{0} + \sin(\alpha) \ket{1} \nonumber\\
 \ket{\varphi_B} &=& \cos(\beta) \ket{0} + \sin(\beta) \ket{1} \nonumber\\
 \ket{\varphi_A} &=& \cos(\gamma) \ket{0} + \sin(\gamma) \ket{1},
\een
and $K_{GHZ}=1/(1+ \cos(\delta)\sin(\delta) \cos(\alpha) \cos(\beta)  \cos(\varphi) )$. The angles belong to the intervals $\delta\in (0,\pi/4]$, $(\alpha,\, \beta,\; \gamma) \in (0,\pi/2]$, and $\varphi \in [0,2\pi)$. Therefore, to generate random GHZ states we sample these angles with a random distribution. 

\subsubsection{$W$-states:} Every $W$ state can be written as

\be
\ket{\psi_\text{W}} = K_{\text{W}} \left( a \ket{001} + b \ket{010} + c \ket{100} + d \ket{000} \right). 
\ee
with $(a,b,c,d)\in(0,1)$, and $K_{\text{W}}=1/\sqrt{a^2+b^2+c^2+d^2}$.  To generate random states we just sample $(a,b,c,d)$ as random numbers from a plain distribution in the interval $(0,1)$.


\begin{thebibliography}{10}

\bibitem{einstein:pr35}
A.~Einstein, B.~Podolsky, and N.~Rosen.
\newblock Can quantum-mechanical description of physical reality be considered
  complete?
\newblock {\em Phys. Rev.}, 47(10):777--780, 1935.

\bibitem{schrodinger:mpcps35}
E.~Schr\"odinger.
\newblock Discussion of probability relations between separated systems.
\newblock {\em Math. Proc. Cam. Phil. Soc.}, 31:555, 1935.

\bibitem{bennet:prl93}
C.~H. Bennett, G.~Brassard, C.~Crepeau, R.~Jozsa, A.~Peres, and W.~K. Wootters.
\newblock Teleporting an unknown quantum state via dual classical and
  einstein-podolsky-rosen channels.
\newblock {\em Phys. Rev. Lett.}, 70:1895, 1993.

\bibitem{pirandola:np15}
S.~Pirandola, J.~Eisert, C.~Weedbrook, A.~Furusawa, and S.~L. Braunstein.
\newblock Advances in quantum teleportation.
\newblock {\em Nature Photonics}, 9:641, 2015.

\bibitem{raussendorf:pra03}
R.~Raussendorf, D.~E. Browne, and H.J. Briegel.
\newblock Measurement-based quantum computation on cluster states.
\newblock {\em Phys. Rev. A}, 68:022312, 2003.

\bibitem{briegel:np09}
H.~J. Briegel, D.~E. Browne, W.~D\"ur, R.~Raussendorf, and M.~Van den Nest.
\newblock Measurement-based quantum computation.
\newblock {\em Nature Physics}, 5(19), 2009.

\bibitem{bennet:prl92}
C.H. Bennett and S.J. Wiesner.
\newblock Communication via one- and two-particle operators on
  einstein-podolsky-rosen states.
\newblock {\em Phys. Rev. Lett.}, 69:2881, 1992.

\bibitem{horodecki:rmp09}
R.~Horodecki, P.~Horodecki, M.~Horodecki, and K.~Horodecki.
\newblock Quantum entanglement.
\newblock {\em Rev. Modern Phys.}, 2009:865--942, 2009.

\bibitem{guhne:pr09}
O.~G\"uhne and G\'eza T\'oth.
\newblock Entanglement detection.
\newblock {\em Phys. Rep.}, 474:1, 2009.

\bibitem{bell:p64}
J.S. Bell.
\newblock On the {E}instein {P}odolsky {R}osen paradox.
\newblock {\em Physics}, 1:195--200, 1964.

\bibitem{peres:prl96}
A.~Peres.
\newblock Separability criterion for density matrices.
\newblock {\em Phys. Rev. Lett.}, 77(8):1413--1415, 1996.

\bibitem{horodecki:pla96}
M.~Horodecki, P.~Horodecki, and R.~Horodecki.
\newblock Separability of mixed states: necessary and sufficient conditions.
\newblock {\em Physics Letters A}, 223(1-2):1 -- 8, 1996.

\bibitem{bru:jmo02}
D.~Bru, J.I. Cirac, P.~Horodecki, F.~Hulpke, B.~Kraus, M.~Lewenstein, and
  A.~Sanpera.
\newblock Reflections upon separability and distillability.
\newblock {\em J. Mod. Opt.}, 49:1399, 2002.

\bibitem{guhne:prl04}
Mohamed Bourennane, Manfred Eibl, Christian Kurtsiefer, Sascha Gaertner, Harald
  Weinfurter, Otfried G\"uhne, Philipp Hyllus, Dagmar Bru\ss{}, Maciej
  Lewenstein, and Anna Sanpera.
\newblock Experimental detection of multipartite entanglement using witness
  operators.
\newblock {\em Phys. Rev. Lett.}, 92:087902, 2004.

\bibitem{plastino:epl09}
A.~R. Plastino, D.~Manzano, and J.~S. Dehesa.
\newblock Separability criteria and entanglement measures for pure states of n
  identical fermions.
\newblock {\em EPL (Europhysics Letters)}, 86(2):20005 (6pp), 2009.

\bibitem{walborn:prl09}
S.~P. Walborn, B.~G. Taketani, A.~Salles, F.~Toscano, and R.~L. de~Matos~Filho.
\newblock Entropic entanglement criteria for continuous variables.
\newblock {\em Phys. Rev. Lett.}, 103:160505, 2009.

\bibitem{dunjko:prl16}
V.~Dunjko, J.M. Taylor, and H.~J. Briegel.
\newblock Quantum-enhanced machine learning.
\newblock {\em Phys. Rev. Lett.}, 117:130501, 2016.

\bibitem{dunjko:rpp18}
V.~Dunjko and H.J. Briegel.
\newblock Machine learning and artificial intelligence in the quantum domain: a
  review of recent progress.
\newblock {\em Rep. Prog. Phys.}, 81:074001, 2018.

\bibitem{manzano:njp09}
D.~Manzano, M.~Paw{\l}owski, and \v{C}. Brukner.
\newblock The speed of quantum and classical learning for performing the k-th
  root of {NOT}.
\newblock {\em New J. Phys.}, 11:113018, 2009.

\bibitem{melnikov:pnas17}
A.~A. Melnikov, H.~P Nautrupa, M.~Krenn, V.~Dunjko, M.~Tiersch, A.~Zeilinger,
  and H.J. Briegel.
\newblock Active learning machine learns to create new quantum experiments.
\newblock {\em Proc. Natl. Acad. Sci.}, 115:1221, 2017.

\bibitem{cao:arxiv17}
Y.~Cao, G.G. Guerreschi, and A.~Aspuru-Guzik.
\newblock Quantum neuron: an elementary building block for machine learning on
  quantum computers.
\newblock ArXiv:1711.11240, 2017.

\bibitem{beer:nc20}
K.~Beer, D.~Bondarenko, Terry Farrelly, T.J. Osborne, R.~Salzmann,
  D.~Scheiermann, and R.~Wolf.
\newblock Training deep quantum neural networks.
\newblock {\em Nat. Comm.}, 11:808, 2020.

\bibitem{torres:njp22}
J.~Torres and D.~Manzano.
\newblock A model of interacting quantum neurons with a dynamic synapse.
\newblock {\em New J. Physics}, 24:073007, 2022.

\bibitem{rosenblatt:pr58}
F.~Rosenblatt.
\newblock The perceptron: A probabilistic model for information storage and
  organization in the brain.
\newblock {\em Psychological Review}, 65(6):386--408, 1958.

\bibitem{minsky_69}
M.L. Minsky.
\newblock {\em Perceptrons}.
\newblock Cambridge, MA:MIT Press, 1969.

\bibitem{roik:pra21}
J.~Roik, K.~Bartkiewicz, A.~Cernoch, and K.~Lemr.
\newblock Accuracy of entanglement detection via artificial neural networks and
  human-designed entanglement witnesses.
\newblock {\em Phys. Rev. App.}, 15(054006), 2021.

\bibitem{asif:sr23}
N.~Asif, U.~Khalid, A.~Khan, T.Q. Duong, and H.~Shin1.
\newblock Entanglement detection with artificial neural networks.
\newblock {\em Scientific Reports}, 13:1562, 2023.

\bibitem{chen:qst22}
Y.~Chen, Y.~Pan, G.~Zhang, and S.~Cheng.
\newblock Detecting quantum entanglement with unsupervised learning.
\newblock {\em Quantum Sci. Technol.}, 7:015005, 2022.

\bibitem{lu:pra18}
S.~Lu, S.~Huang, K.~Li, J.~Li, J.~Chen, D.~Lu, Z.~Ji~Y. Shen, D.~Zhou, and
  B.~Zeng.
\newblock Separability-entanglement classifier via machine learning.
\newblock {\em Phys. Rev. A}, 98:012315, 2018.

\bibitem{gao:prl18}
J.~Gao, L.-F. Qiao, Z.-Q. Jiao, Y.-C. Ma, C.-Q. Hu, R.-J. Ren, A.-L. Yang,
  H.~Tang, M.-H. Yung, and X.-M. Jin.
\newblock Experimental machine learning of quantum states.
\newblock {\em Phys. Rev. Lett.}, 120:240501, 2018.

\bibitem{girardin:prr22}
N.~Brunner A.~Girardin and T.~Krivachy.
\newblock Building separable approximations for quantum states via neural
  networks.
\newblock {\em Phys. Rev. Res.}, 4:023238, 2022.

\bibitem{mcculloch:bmb43}
W.~Mcculloch and W.~Pitts.
\newblock A logical calculus of ideas immanent in nervous activity.
\newblock {\em Bulletin of Mathematical Biophysics}, 5:127, 1943.

\bibitem{russel_03}
S.J. Russel and P.~Norvig.
\newblock {\em Artificial intelligence - {A} modern approach. Second edition}.
\newblock Prentice Hall, New Jersey, 2003.

\bibitem{nair:proc10}
V.~Nair and G.E. Hinton.
\newblock Rectified linear units improve restricted boltzmann machines.
\newblock In {\em Proceedings of the 27th International Conference on Machine
  Learning (ICML)}, 2010.

\bibitem{du:m22}
K.-L. Du, C.-S. Leung, W.~H. Mow, and M.~N.~S. Swamy.
\newblock Perceptron: Learning, generalization, model selection, fault
  tolerance, and role in the deep learning era.
\newblock {\em Mathematics}, 10:4730, 2022.

\bibitem{diederik:proc15}
P.K. Diederik and J.~Ba.
\newblock Adam: A method for stochastic optimization.
\newblock In {\em Proceedings of the 3rd International Conference for Learning
  Representations}, 2015.

\bibitem{riedmiller:proc93}
M.~Riedmiller and H.~Braun.
\newblock A direct adaptive method for faster backpropagation learning: The
  rprop algorithm.
\newblock In {\em IEEE International Conference on Neural Networks}, 1993.

\bibitem{glorot:proc10}
X.~Glorot and Y.~Bengio.
\newblock Understanding the difficulty of training deep feedforward neural
  networks.
\newblock In {\em Proceedings of the Thirteenth International Conference on
  Artificial Intelligence and Statistics}, volume~9, page 249, 2010.

\bibitem{zyczkowski:pra98}
K.~Zyczkowski, P.~Horodecki, A.~Sanpera, and M.~Lewenstein.
\newblock Volume of the set of separable states.
\newblock {\em Phys. Rev. A}, 58:883, 1998.

\bibitem{werner:pra89}
Reinhard~F. Werner.
\newblock Quantum states with {Einstein-Podolsky-Rosen} correlations admitting
  a hidden-variable model.
\newblock {\em Phys. Rev. A}, 40(8):4277--4281, 1989.

\bibitem{greenberger_89}
D.~M. Greenberger, M.~Horne, and A.~Zeilinger.
\newblock {\em Bell's Theorem, Quantum Theory, and Conceptions of the
  Universe}, page~69.
\newblock Kluwer Acad. Publ., Dordrecht, 1989.

\bibitem{dur:pra00}
W.~D\"ur, G.~Vidal, and J.~I. Cirac.
\newblock Three qubits can be entangled in two inequivalent ways.
\newblock {\em Phys. Rev. A}, 62:062314, 2000.

\bibitem{nielsen_00}
M.A. Nielsen and I.L. Chuang.
\newblock {\em Quantum Computation and Quantum Information}.
\newblock Cambridge Univ. Press, Cambridge, 2000.

\end{thebibliography}
\end{document}